\documentclass[aps,preprint,groupedaddress,showpacs]{revtex4}
\begin{document}

\title{Polarized far-infrared and
Raman spectra of SrCuO$_2$ single crystals }
\author{Z. V. Popovi\'c$^{a,c}$\footnote{Present address: Laboratorium voor Vaste-Stoffysica en Magnetisme, K. U. Leuven,
Celestijnenlaan 200D, B-3001 Leuven, Belgium, Fax:+32-16-327983,
e-mail:Zoran.Popovic@fys.kuleuven.ac.be}, M. J.
Konstantinovi\'c$^{b,c}$, R. Gaji\'c$^{c}$,
  C. Thomsen$^{d}$, U. Kuhlmann$^{d}$ and A. Vietkin$^{e}$ }

\affiliation{ $^a$ Laboratorium voor Vaste-Stoffysica en
Magnetisme, K. U. Leuven, Celestijnenlaan 200D, B-3001 Leuven,
Belgium } \affiliation{$^b$ Max-Planck-Institut f\"ur
    Festk\"orperforschung, Heisenbergstrasse 1, D-70569 Stuttgart, F. R. Germany}
\affiliation{$^c$ Institute of Physics, P. O. Box 68, 11080
Belgrade, Yugoslavia} \affiliation {$^d$ Institut f\"ur
    Festk\"orperphysik, Technische Universit\"{a}t Berlin, Hardenbergstrasse 36, D-10623 Berlin F. R.
Germany} \affiliation{$^e$ Physics Department, Moscow State
University, 119899 Moscow, Russia}

\begin{abstract}
We measured polarized far-infrared reflectivity and Raman scattering spectra of
SrCuO$_2$ single crystals. The frequencies for infrared-active modes were
determined using an oscillator-fitting procedure of reflectivity data. The
Raman spectra were measured at different temperatures using several laser
energies $\omega_L$. In addition to eight of twelve Raman active modes,
predicted by factor-group analysis, we observed a complex structure in the
Raman spectra for polarization parallel to the {\bf c}-axis, which consists of
Raman-allowed A$_g$ symmetry modes, and B$_{1u}$ LO infrared-active
(Raman-forbidden) modes of the first and higher order as well as their
combinations. The Raman-forbidden modes have a stronger intensity at higher
$\omega_L$ than the Raman-allowed ones. In order to explain this resonance
effect, we measured the dielectric function and optical reflection spectra of
SrCuO$_2$ in the visible range. We show that the Raman-allowed A$_g$ symmetry
modes are resonantly enhanced when a laser energy is close to $E_0$, while
Raman-forbidden (IR-active) modes resonate strongly for laser line energies
close to the electronic transition of higher energy gaps.
\end{abstract}
\pacs{ 78.30.Hv; 78.20.Ci; 63.20.Dj; 71.27.+a;} \maketitle

\section{Introduction}
Strontium copper oxide belongs to the new family of 1D antiferromagnetic
insulators the properties of which have been subject of intensive studies in
recent years \cite {a1, a2, a3, a4, a5, a6, a7, a8}. This oxide, grown in
single crystalline form under ambient conditions, has an orthorhombic unit cell
(space group $D_{2h}^{17}$ - $Cmcm$) with parameters a=0.3577 nm, b=1.6342 nm,
c= 0.39182 nm and Z=4 \cite {a9, a10}. The compound has a unique structure
consisting of CuO$_4$ squares, mutually connected via common edges that form
double copper chains (Fig. 1). These double chains, stretched along the {\bf
c}-axis, could be taken as a quasi-one-dimensional antiferromagnetic chains of
Cu$^{2+}$ ions with S=1/2, since the 90$^o$ Cu-O-Cu superexchange interactions
between neighboring chains is negligible in comparison with the 180$^o$ Cu-O-Cu
superexchange interactions along the chains \cite {a2}. Magnetic susceptibility
measurements of SrCuO$_2$ show that the Cu$^{2+}$ moments order
antiferromagnetically at about 2 K \cite {a4}. The exchange interaction energy
J is estimated to be $2100 \pm 200$ K \cite {a5}.

The Raman scattering spectra of SrCuO$_2$ were reported by Ref. \cite {a6}. The
A$_g$ symmetry modes at 188, 262, 543 and 558 cm$^{-1}$ were clearly resolved
at room temperature. The Raman spectra for the polarization parallel to Cu-O
chains show additional structures ascribed to either phonons strongly coupled
to spinons or pairs of unbound spinons that resonate as the excitation energy
approaches the charge-transfer (CT) gap \cite {a6}.

More recently \cite {a7, a8}, micro-Raman and the far-infrared
(FIR) measurements of polycrystalline SrCuO$_2$ were reported. The
assignment of the observed modes was given according to a shell
model lattice-dynamics calculation \cite {a7}. In addition to the
Raman-allowed modes, numerous lines with A$_g$ symmetry appear for
polarization parallel to the Cu-O chains only. These modes
originate from the B$_{1u}$ LO modes and their overtones and
combinations, becoming Raman-active via Fr\"{o}hlich interaction
\cite {a8}. It was also shown in Ref. \cite {a8}, using the
calculated combination two-phonon density of states of only
B$_{1u}$ dispersion curves, that non-phonon excitations \cite {a6}
are not necessary for the understanding of Raman spectra on
SrCuO$_2$. Infrared and Raman unpolarized spectra of the
tetragonal modification of this oxide have been analyzed in
\cite{a11}. Some of our preliminary results we discussed in
\cite{a12}.

In this paper we report measurements of the far-infrared reflectivity and Raman
scattering spectra of single crystal SrCuO$_2$, which enables us to make full
use of polarization selection rules. The TO and LO frequencies of B$_{1u}$,
B$_{2u}$ and B$_{3u}$ infrared-active modes were obtained by applying an
oscillator-fitting procedure to the reflectivity data. The Raman spectra were
measured at 10 K and at room temperature. Eight Raman active modes of different
symmetry were resolved. The Raman spectra for polarization along the Cu-O
chains show a rich structure consisting of Raman-allowed A$_g$ symmetry modes
and B$_{1u}$ LO infrared-active (Raman-forbidden) modes of the first and higher
order as well as their combinations. These modes become more pronounced when Ar
laser lines of higher photon energy are used. In order to explain this resonant
effect, we analyzed optical properties of this oxide in the visible range on
the basis of optical reflection and dielectric function measurements.

\section{Experimental details}

The present work was performed using single crystal sample with a
size of about 20, 3 and 7 mm along the a, b and c axes,
respectively. The Raman spectra were measured in the
backscattering configuration using Ar and Kr lasers for excitation
sources. For low temperature measurements we used a closed-cycle
cryostat. The scattered light was detected by a Jobin Yvon U-1000
(XY-800 Dilor) double (triple) monochromator with 1800 groove/mm
gratings and a RCA 31034 A photomultiplier (CCD) as a detector.
The far-infrared reflectivity measurements were carried out with a
BOMEM DA-8 FIR spectrometer. A DTGS pyroelectric detector was used
to cover the wave number region from 100 to 700 cm$^{-1}$. The
spectra were collected with the 2 cm$^{-1}$ resolution.  We
measured the pseudodielectric function with a help of a
rotating-analyzer ellipsometer. A Xe-lamp was used as a light
source, a double monochromator with 1200 lines/mm gratings and an
S20 photomultiplier tube as a detector. The polarizer and analyzer
were Rochon prisms.  The measurements were performed in the
1.2-5.6 eV energy range. Optical reflectivity was measured in the
200-2500 nm spectral range using Perkin - Elmer, model Lambda
spectrophotometer.

\section{Results and Discussion}
Room temperature polarized far-infrared reflectivity spectra of SrCuO$_2$, in
the spectral range from 100 to 675 cm$^{-1}$, are shown in Fig. 2. The circles
are the experimental data and the solid lines represent the spectra computed
using a four-parameter model for the dielectric constant:

\begin{equation}
\epsilon(\omega)=\epsilon_{\infty} \prod_{j=1}^{n}
\frac{\omega_{LO,j}^2-\omega^2+\imath
\gamma_{LO,j}\omega}{\omega_{TO,j}^2-\omega^2+\imath\gamma_{TO,j}\omega},
\label{1}
\end{equation}

where $\omega_{LO,j}$ and $\omega_{TO,j}$ are longitudinal and transverse
frequencies of the j$^{th}$ oscillator, $\gamma_{LO,j}$ and $\gamma_{TO,j}$ are
their corresponding dampings, and $\epsilon_{\infty}$ is the high-frequency
dielectric constant. The best-oscillator-fit parameters are listed in Table I.
The static dielectric constant, given in Table I, was obtained using the
generalized Lyddane-Sachs-Teller relation:

\begin{equation}
\epsilon_0=\epsilon_{\infty} \prod_{j=1}^{n}
\frac{\omega_{LO,j}^2}{\omega_{TO,j}^2}. \label{2}
\end{equation}
As can be seen in Fig. 2, the agreement between the observed and the calculated
reflectivity spectra is satisfactory. Deviations between these spectra occurs
only there where the leakage from another polarization exists.

The factor-group analysis for this crystal predicts:
\begin{equation}
\Gamma_{SrCuO_2}= 4A_g (aa, bb, cc) + 4B_{1g}(ab) + 4B_{3g}(bc) +
3B_{1u}({\bf E}||{\bf c}) + 3B_{2u}({\bf E}||{\bf b}) +
3B_{3u}({\bf E}||{\bf a}) \label{3}
\end{equation}

According to Fig. 2 and Table I, all nine infrared active modes are observed.
Comparing our single crystal reflectivity spectra with those previously
published for polycrystalline samples (Table I), we found a difference between
the TO(LO) frequencies which, for higher energy modes, reaches a value of 10
(25) cm$^{-1}$. Such a big difference may be expected because the reflectivity
measurements carried out on polycrystalline samples give good results for TO
and LO mode frequencies in the case of isolated oscillators only. Here we have
different symmetry oscillators with almost identical TO and different LO
frequencies. It is not possible, therefore, to extract precise values of the
optical parameters either by Kramers-Kronig analysis or oscillator-fitting
procedure from the unpolarized reflectivity spectra (see inset in Fig. 2).

We also measured the unpolarized reflectivity spectra at low temperature of 10
K. Note the strong frequency shift of the LO$_3$ mode from 586 cm$^{-1}$ to 596
cm$^{-1}$ upon lowering the temperature to 10 K (dashed curve in inset in Fig.
2). No evidence of any additional structure was observed.

The polarized Raman scattering spectra of SrCuO$_2$ measured at room
temperature are shown in Fig. 3. According to the assignments given in Ref.
\cite {a7}, the lowest frequency A$_g$ mode at 186 cm$^{-1}$ originates from
the vibration of Sr atoms; the mode at 262 cm$^{-1}$ represents Cu vibrations,
while the other two A$_g$ high frequency modes are assigned to O$^{Sr}$ and
O$^{Cu}$ vibrations, respectively. Let us note that the polarization dependence
of the strength of the A$_g$ modes, shown for blue excitation at 488 nm in Fig.
3, disappears for the red excitation of the Kr laser at 647.1 nm. Namely, the
A$_g$ modes at 186 and 262 cm$^{-1}$ are nearly of the same intensity for the
647.1 nm excitation at 300 K.

The Raman spectra for crossed polarization are shown in the Inset of Fig. 3.
The B$_{1g}$ symmetry modes can be seen for (xy) polarization. For this
polarization we clearly observed the Cu vibration mode at about 150 cm$^{-1}$
and O$^{Sr}$ vibration mode at 229 cm$^{-1}$. We observed also a wide structure
at about 215 cm$^{-1}$ which represents a superposition of two peaks at 211 and
217 cm$^{-1}$. The peak at 217 cm$^{-1}$ is the mode of B$_{3g}$ symmetry. Thus
we identified three of four B$_{1g}$ modes. The fourth mode of this symmetry
was observed at 78 cm$^{-1}$ in \cite {a7}. Very strong rotational modes of
nitrogen from air do not allow us to resolve this mode in our experiment here.
For (zy) polarization we found only one mode at about 217 cm$^{-1}$, which
represents Cu vibrations \cite {a7}. Other modes of this symmetry are probably
of very week intensity.

The Raman spectra of SrCuO$_2$ for the (cc) polarization
configuration (incident and scattered light are polarized along
the Cu-O chains) at 10K are shown in Fig. 4. These spectra consist
of Raman-allowed A$_g$ symmetry modes and
Raman-forbidden-infrared-active B$_{1u}$ LO modes, their overtones
and combinations, becoming Raman-active via Fr\"{o}hlich
interaction \cite {a8}. Appearance of the infrared active modes in
the Raman spectra of CuO based materials is observed when the
laser energy is close to the charge-transfer (CT) gap transition
\cite {a13,a14,a15,a16}. It was shown \cite {a14} that, depending
on coordination, the values of the CT gap decrease from 2.0 eV in
the case of octahedral CuO$_2$ layers in La$_2$CuO$_4$ to 1.5 eV
in square-type CuO$_2$ sheets in Nd$_2$CuO$_4$. To best of our
knowledge there is no experimental data about energy gap of
SrCuO$_2$. We attempted to extract some information from optical
conductivity spectra of the Ca$_{0.85}$Sr$_{0.15}$CuO$_2$ crystal.
We have obtained the main conductivity peak at 1.6 eV and a less
prominent tail at 2.1 eV by deconvolution of the optical
conductivity spectra of Ca$_{0.85}$Sr$_{0.15}$CuO$_2$ given in
Fig. 1(e) of Ref. \cite {a16}, who obtained their conductivity
spectra by Kramers-Kronig analysis of the reflectivity data,
measured at room temperature with the polarization parallel to the
Cu-O chains. Since the substitution of Ca for Sr atoms produces
crystal structure transition from orthorhombic to tetragonal and a
corresponding change of electronic structure, it was not possible
to use these data as a gap values of SrCuO$_2$. Instead, we
measured the dielectric function and optical reflectivity of
SrCuO$_2$ in the visible and UV spectral range. The dielectric
function $\epsilon_2$ of SrCuO$_2$ is shown in Fig. 5. These
spectra were computed from the measured Fourier coefficients using
the equations for an isotropic case. Consequently, $\epsilon_2$
represents a complicated average of the projections of the
dielectric tensor on the sample surface. Bands with the energies
of 3.15, 4.1 and 5.23 eV were found for the {\bf c} axis, and at
about 1.74, 2.0, 4.02, and 5.1 eV for the {\bf a}-axis in the
plane of incidence, respectively. By comparison of ellipsometric
data in Fig. 5 with optical reflectivity spectra shown in the
right Inset of Fig. 5, we concluded that dielectric functions
coincide with reflectivity spectra in the spectral range from 1.6
to 5.5 eV. In order to check from which polarization originate the
lowest energy maxima in optical reflectivity spectra at about 1.5
eV we measured dielectric functions in 1.2 - 1.8 eV range, also
(left Inset of Fig. 5). We found that only for $E
\parallel c$ polarization exists a weak structure at about 1.4 eV. The
position of this line is obtained by a Lorentzian deconvolution.
We conclude that the lowest energy gap in SrCuO$_2$ (E$_0$) is at
about 1.4 eV (1.74 eV) for $E
\parallel c$ $(E
\parallel a)$ polarization.

According to the spectra in Fig. 4, the enhancement of the Raman-forbidden
modes appears at higher energies than the gap value. Namely, in the case of the
647.1 nm excitation, Raman-allowed Ag modes at 190, 265 and 544 cm$^{-1}$ are
the dominant structures. With an increase in excitation energy to 514.5 nm,
additional structures enhance resonantly and new modes, superimposed on a wide
background, appear. The frequencies and the assignment of all observed modes
are given in Table II. A more detailed analysis indicates that the
Raman-allowed A$_g$ modes in combination with the LO modes also contribute to
the two-phonon spectra. For the modes at 796 cm$^{-1}$ (A) and 1168 cm$^{-1}$
(B) no corresponding combination of modes has been found, and we presume that
they originate from zone-edge phonons.

The highest intensity mode at about 1195 cm$^{-1}$ is asymmetric
due to superposition of several modes. We fitted this structure
with four Lorentzians in order to determine frequencies at which
maxima occur (upper inset in Fig. 4). For the 514.5 nm line, the
dominant contribution to this mode comes from 2LO$_3$. Using the
647.1 nm excitation, this mode becomes weaker than the mode at
about 1170 cm$^{-1}$ (B), giving the impression that its frequency
decreases by lowering the excitation energy.

Having above in mind, we conclude that the 2LO overtone processes
are not resonantly enhanced with a laser line energy (1.91 eV)
near the lowest energy gap but with a laser energy close to the
energies at which the light absorption is significantly increased
(light absorption maxima for higher energy transitions). This
effect, observed for the first time in II-IV Mg$_2$X (X=Si, Ge,
Sn) type semiconductors \cite {a17,a18,a19}, is attributed to
spatial dispersion effects arising from the finite wave-vectors of
phonons and/or the incident and scattered electromagnetic
radiation. As it is shown also in \cite {a19}, the strength of
resonance in Mg$_2$X at E$_0$ is weaker than that at higher energy
transitions, which is explained by the weaker electron-phonon
interaction at E$_0$.

In 1D antiferromagnetic  S=1/2 systems, the excitation spectrum
consists of magnon (more precisely spinon) continuum that builds
up at nonzero momentum (largest contribution is at zone boundary).
Therefore, some contribution in form of two-spinon continuum might
be expected in the Raman spectra. However, in the Raman as well as
in the infrared spectra of SrCuO$_2$, measured up to 6000
cm$^{-1}$ ($J\sim1500 cm^{-1}$), we did not find any feature which
could be assigned as magnetic in origin, as in the case of
Sr$_2$CuO$_3$ \cite{a20}. The reason for this is probably very
small spectral weight of these excitations. Besides, the
two-spinon Raman intensity, in perfect 1D case, is expected to be
largest at small energies and continuously decreases towards
higher energies. There, at low energies, we see strong two-phonon
continuum that mask week two-spinon continuum. At the other hand,
the magnetic structure of SrCuO$_2$ is of zigzag spin-chain type
\cite {a21,a22}, different to that of Sr$_2$CuO$_3$, which also
might be the reason for the different Raman and IR spectra in
these two compounds.

In conclusion, we presented the polarized far-infrared
reflectivity, Raman scattering spectra and ellipsometric as well
as optical reflectivity results of single crystal SrCuO$_2$. In
addition to Raman-allowed modes, the Raman-forbidden -
infrared-allowed B$_{1u}$ LO modes are observed for polarization
parallel to the c-axis. These modes are resonantly enhanced for a
laser energy close to the electronic transition of higher energy
gaps. This effect could be either explained by the
exciton-mediated multi-phonon Raman scattering or attributed to
the spatial dispersion effects arising from the finite wave
vectors of the phonons and/or the incident and scattered
electromagnetic radiation. Resonant Raman scattering in charge
transfer 2D insulating cuprates \cite {a15, a23} shows that
IR-modes and their overtones resonate much more strongly near the
optical absorption edge. In SrCuO$_2$ (1D cuprate) IR modes
resonate at energies higher than fundamental gap. It means that
there are significant differences between the electronic structure
of SrCuO$_2$ and other 2D cuprates.

\section*{Acknowledgment}

We thank Prof. E. Anastassakis for helpful discussion. Z.V.P. acknowledges
support from Alexander von Humboldt Stifting-Bonn and from the Research Council
of the K.U. Leuven and DWTC.  The work at the K.U. Leuven is supported by the
Belgian IUAP and Flemish FWO and GOA Programs. M.J.K thanks Roman Herzog - AvH
for partial financial support.

\clearpage

\begin{figure}
\caption {Schematic representation of the SrCuO$_2$ crystal
structure.} \label{fig1}
\end{figure}

\begin{figure}
\caption {Room temperature polarized far-infrared reflectivity
spectra of SrCuO$_2$ single crystal in the spectral range 100-675
cm$^{-1}$ for all three principal polarizations. The experimental
values are given by open circles. The solid lines represent the
calculated spectra obtained by a fitting procedure described in
text. Inset: Unpolarized reflectivity spectra of SrCuO$_2$ single
crystal at room temperature (solid line) and 10 K (dashed line).}
\label{fig2}
\end{figure}

\begin{figure}
\caption {Raman scattering spectra of SrCuO$_2$ at room temperature for
different polarization configurations. Inset: B$_{1g}$ and B$_{3g}$ scattering
configuration; the A$_g$ modes appear through leakage.} \label{fig3}
\end{figure}

\begin{figure}
\caption {The (cc) polarized Raman spectra of SrCuO$_2$ at 10 K
for (a) 647.1 nm and (b) 514.5 nm excitation. Inset: Raman spectra
measured with 647.1 nm and 514.5 nm lines at 10 K in the spectral
range 1000-1400 cm$^{-1}$, deconvoluted by Lorentzians. }
\label{fig4}
\end{figure}

\begin{figure}
\caption {Room temperature imaginary part ($\epsilon_2$) of the
pseudodielectric function of SrCuO$_2$. The spectra of the (010)
surface taken with a-axis ($E \parallel a$) and c-axis ($E
\parallel c$), parallel to the plane of incidence. Inset left: The
$\epsilon_2(\omega)$ spectra in the 1.2 - 1.8 eV range for $E
\parallel c$ polarization. Inset right: Unpolarized room temperature
reflectivity spectra of SrCuO$_2$ in the 1 - 5.5 eV range. }
\label{fig5}
\end{figure}
\begin{table}
\caption{Frequencies and dampings (in cm$^{-1}$) of infrared and Raman active
modes in SrCuO$_2$.}
\begin{tabular}{ccccccccc}
\hline \hline

Polarization & $\omega_{TO}$ &Ref.[7]& $\gamma_{TO}$ &
$\omega_{LO}$ &Ref.[7]& $\gamma_{LO}$ &
$\varepsilon_\infty$&$\varepsilon_0$
\\ \tableline & 142.5 & (148) & 5  & 157 & (154) & 4 \\  ${\rm\bf
E\parallel c}$ & 307 & (291) & 13 & 345 & (370) & 25 & 4 & 7.7
\\ & 520 & (506) & 30 & 582 & (561) & 30 \\\tableline & 200 & - & 15 & 221 & - & 15 \\${\rm\bf
E\parallel b}$ & 506 & - & 50 & 560 & - & 40 & 2 & 3.7
\\ & 580 &
(577) & 40 & 645 & (527) & 70 \\ \tableline & 143 & (137) & 2 &
145 & (146) & 2
\\ ${\rm\bf E\parallel a}$ & 178 & (172) & 12 & 213 & (218) & 15 &
2.5 & 6.5 \\ & 300 & 25 & 400 & - & 40 \\ \tableline $\bf A_g:$ &
186 & (188) & $\bf B_{1g}:$ & - & 78 & $\bf B_{3g}:$ & - & (120)
\\ & 262 & (263) & {} & 150 & (150) & {} & 217 & (219) \\ & 543 & (544)& {} & 211 & - & {} & - & (320)
\\& 558 & (560) & {} & 229 & (231) & {} & {-} & - \\
\hline \hline
\end{tabular}
\end{table}

\begin{table}
\caption{Frequencies (in cm$^{-1}$) and assignment of modes observed for (cc)
polarization at T=10 K.}
\begin{tabular}{ccc}
\hline\hline No of peaks& Frequency & Remark\\\tableline

1&156&LO$_1$\\2&190&A$_{g1}$\\3&265&A$_{g2}$\\4&347&LO$_2$\\
5&455&A$_{g1}$+A$_{g2}$\\6&503&LO$_1$+LO$_2$\\7&530&2A$_{g2}$\\
8&544&A$_{g3}$\\9&562&A$_{g4}$\\10&596&LO$_3$\\11&610&A$_{g2}$+LO$_2$\\12&694&2LO$_2$\\
13&733&A$_{g1}$+A$_{g3}$\\14&755&LO$_{1}$+LO$_{3}$\\15&796&A\\16&941&LO$_{2}$+LO$_{3}$\\17&1040&2LO$_{3}$-LO$_{1}$\\
18&1130&A$_{g2}$+LO$_3$\\19&1168&B\\20&1195&2LO$_3$\\\hline\hline
\end{tabular}
\end{table}

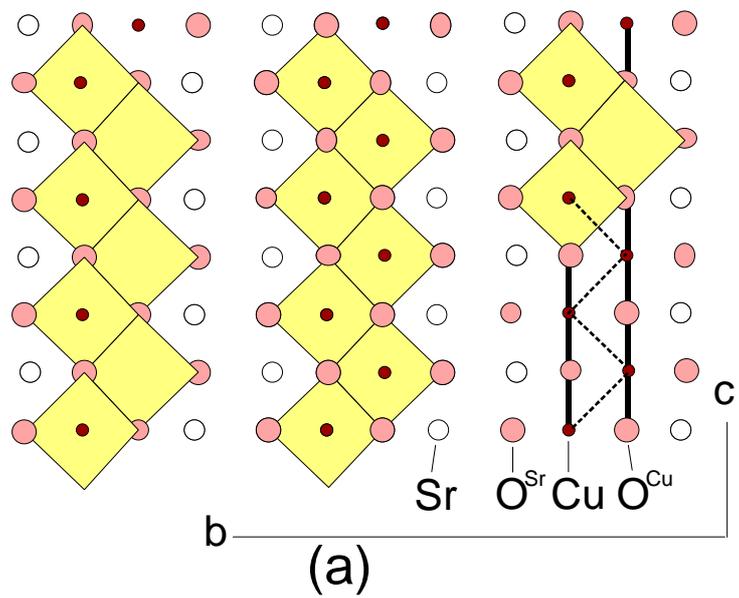
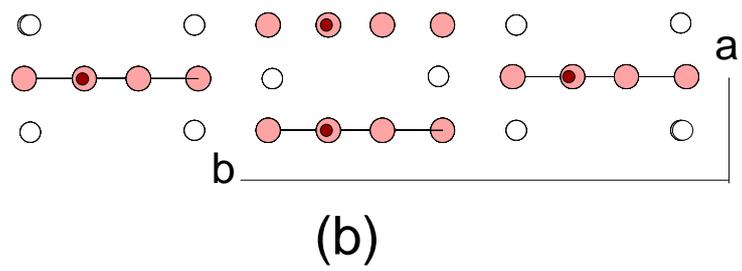
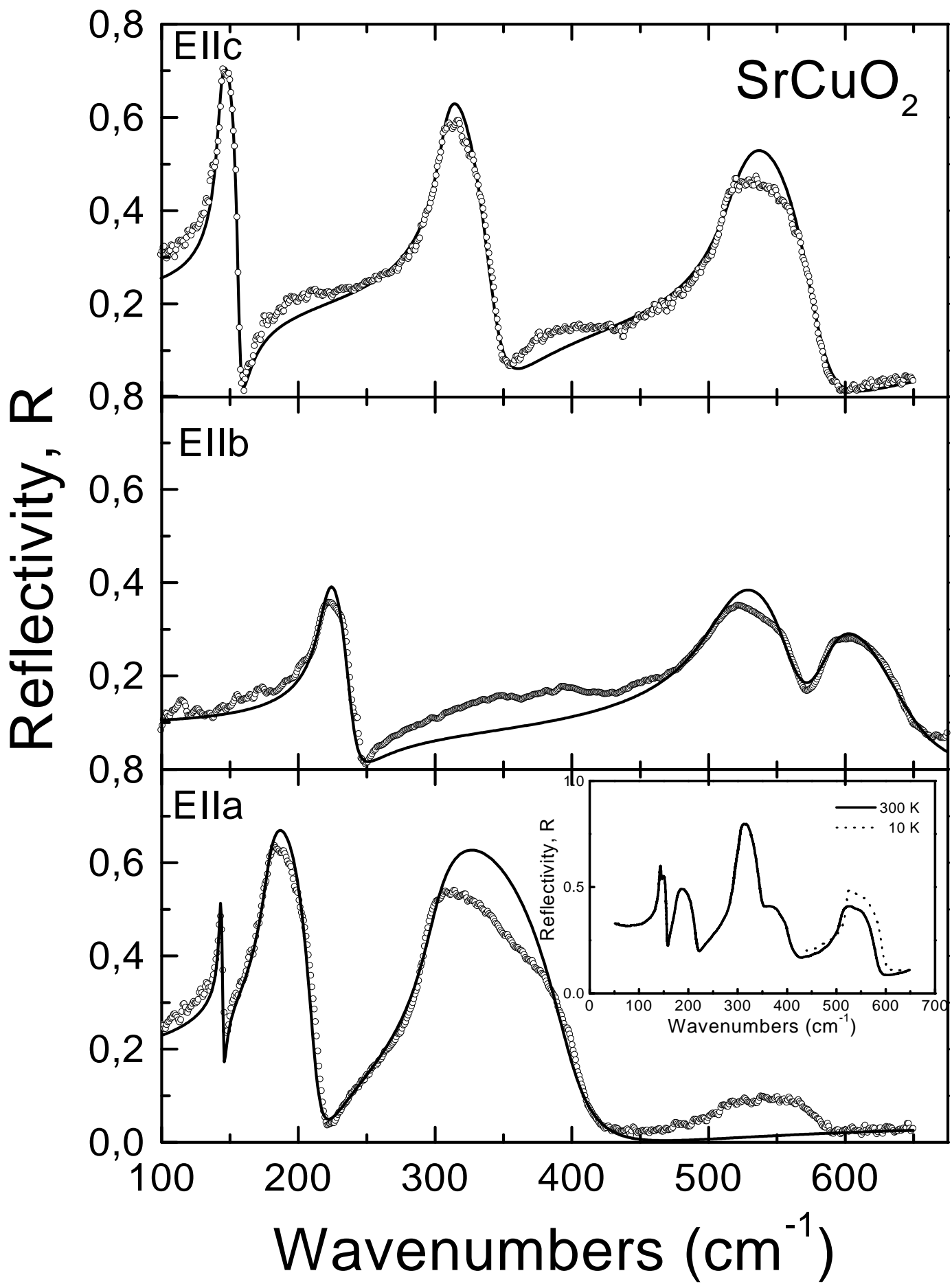
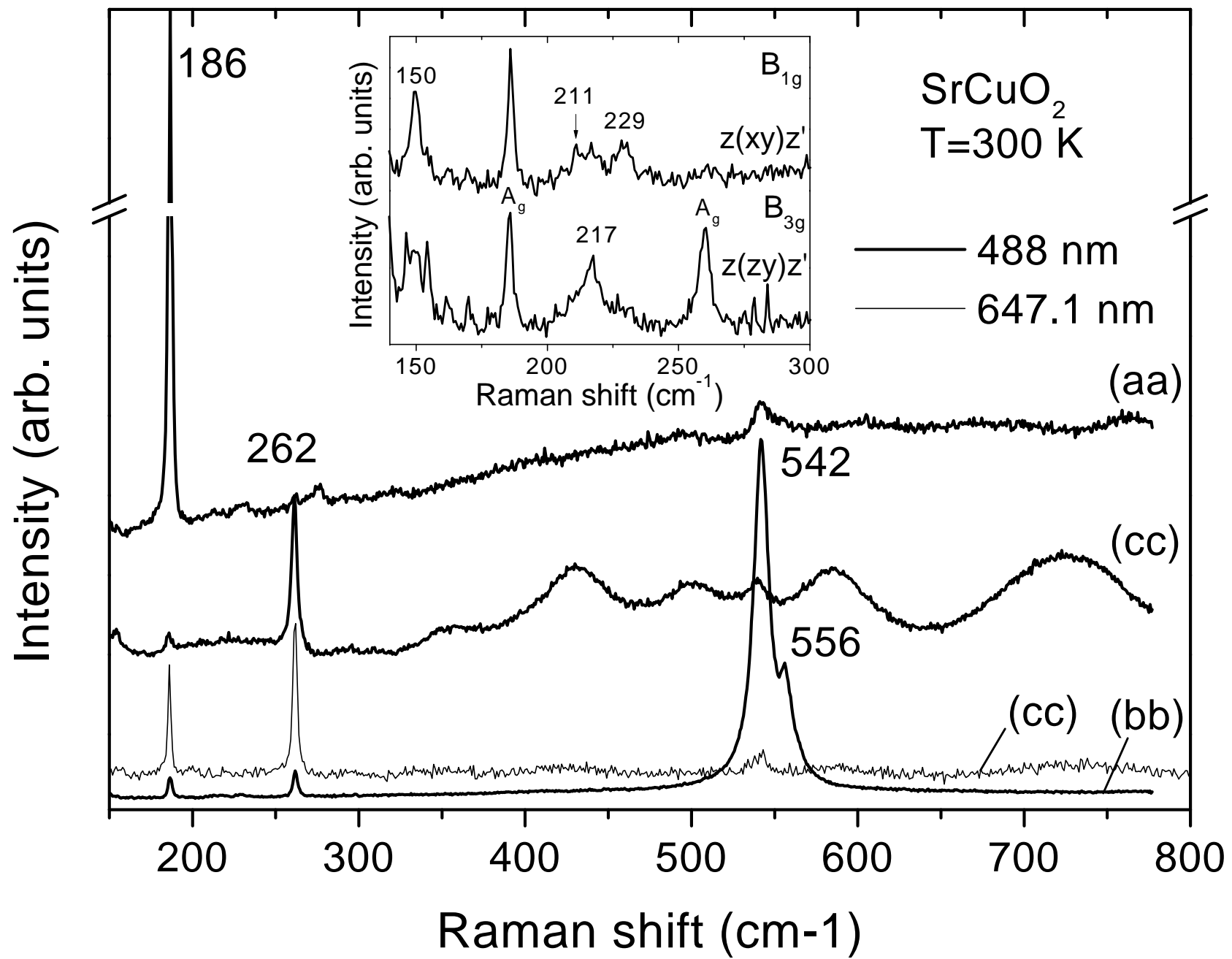
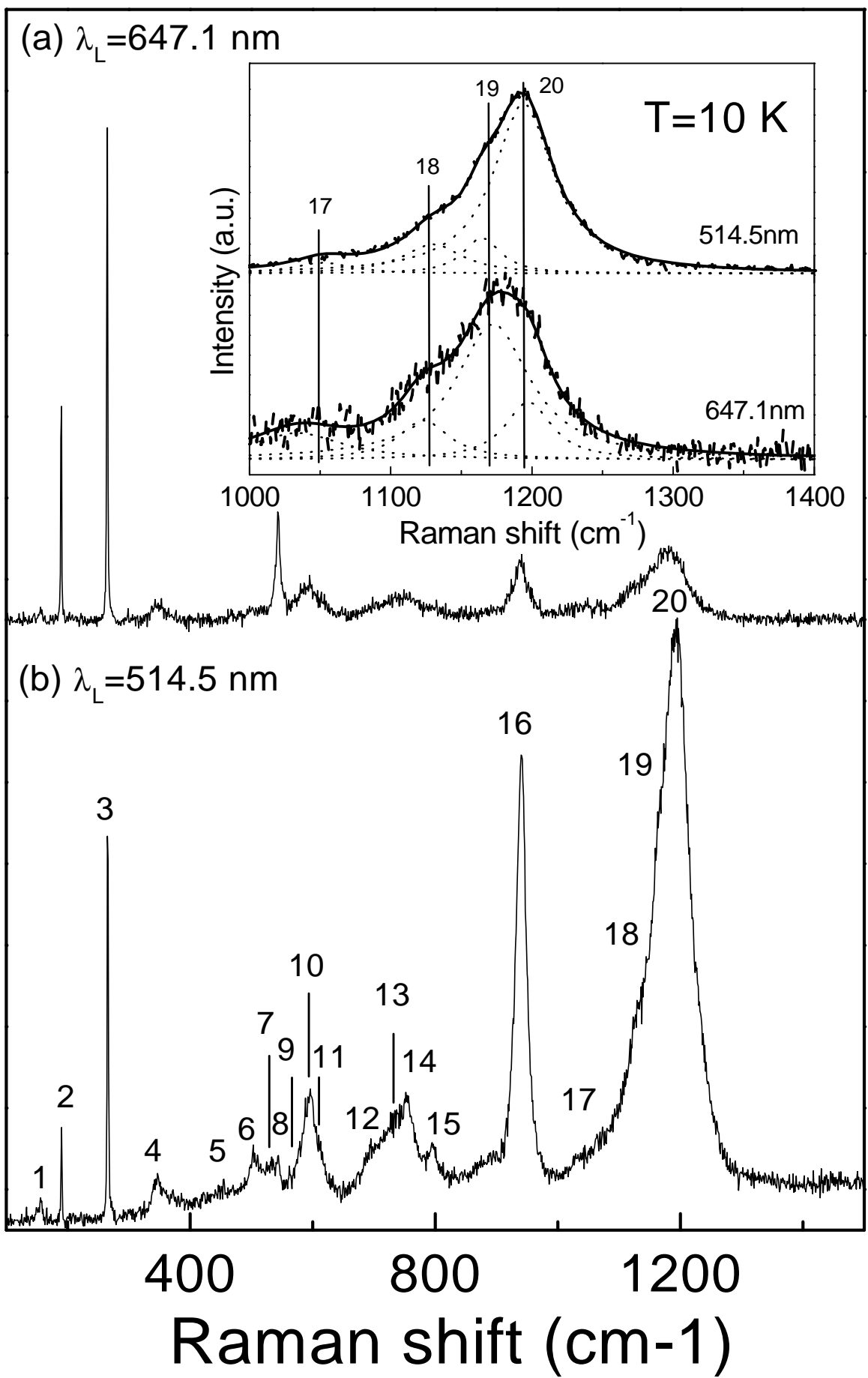
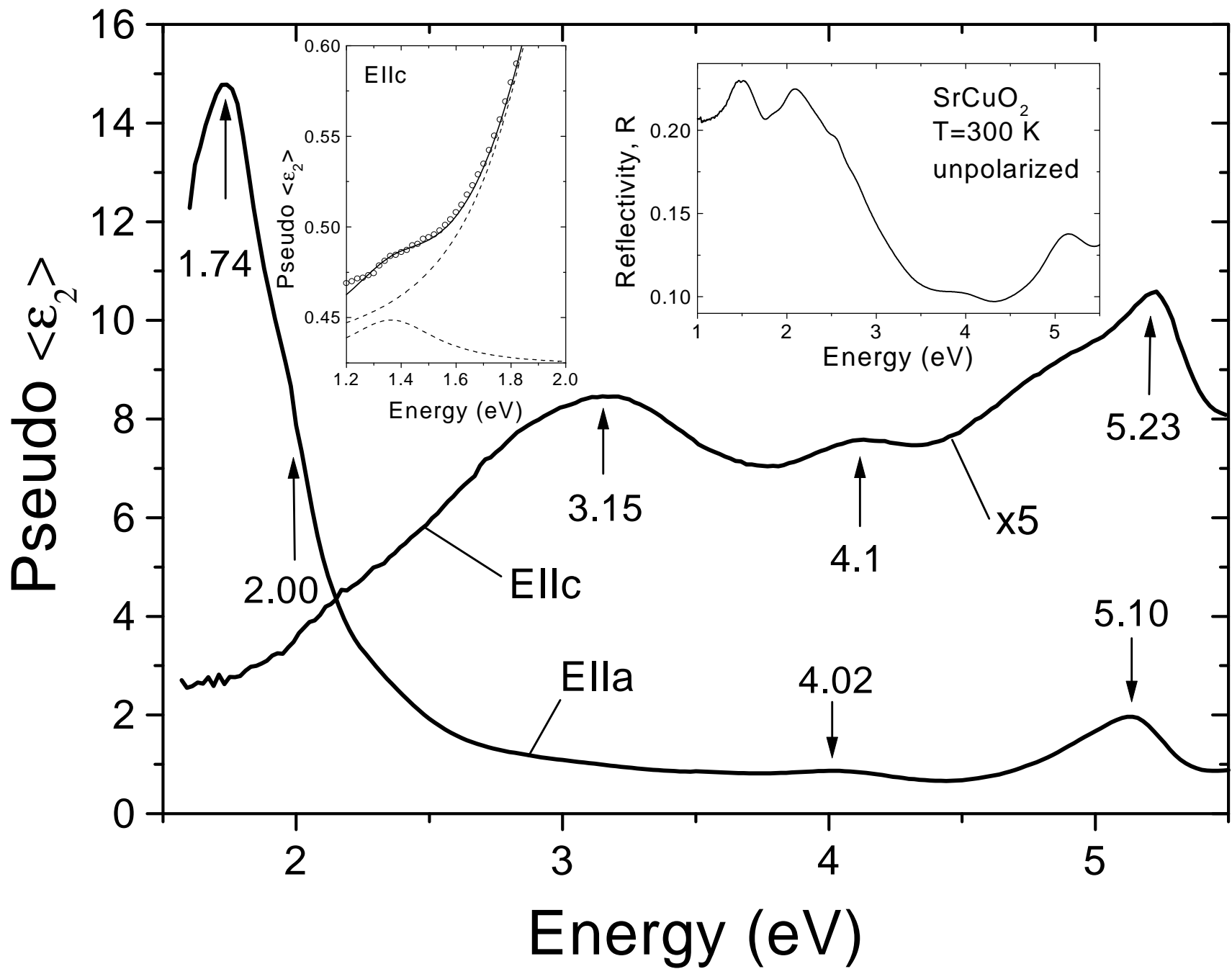
\end{document}